\documentclass[12pt,preprint]{aastex}
\usepackage{natbib}
\usepackage{color}
\begin{document}
\title{ON THE 2012 OCTOBER 23 CIRCULAR RIBBON FLARE: EMISSION FEATURES AND MAGNETIC TOPOLOGY}
\author{\sc{Kai Yang$^{1,2}$, Yang Guo$^{1,2}$, M. D. Ding$^{1,2}$}}
\affil{$^1$ School of Astronomy and Space Science, Nanjing University, Nanjing 210023, China} \email{guoyang@nju.edu.cn, dmd@nju.edu.cn}
\affil{$^2$ Key Laboratory for Modern Astronomy and Astrophysics (Nanjing University), Ministry of Education, Nanjing 210023, China}
\begin{abstract}
Circular ribbon flares are usually related to spine-fan type magnetic topology containing null-points.
In this paper, we investigate an X-class circular ribbon flare on 2012 October 23, using the multi-wavelength data from the \textit{Solar Dynamics Observatory}, \textit{Hinode}, and the \textit{Ramaty High Energy Solar Spectroscopic Imager}.
In \ion{Ca}{2} H emission, the flare showed three ribbons with two highly elongated ones inside and outside a quasi-circular one, respectively.
A hot channel was displayed in the extreme ultraviolet (EUV) emissions that infers the existence of a magnetic flux rope.
Two hard X-ray (HXR) sources in the 12--25 keV energy band were located at the footpoints of this hot channel.
Using a nonlinear force-free magnetic field extrapolation, we identify three topological structures: (1) a 3D null-point, (2) a flux rope below the fan of the null-point, and (3) a large-scale quasi-separatrix layers (QSL) induced by the quadrupolar-like
magnetic field of the active region.
We find that the null-point is embedded within the large-scale QSL.
In our case, all three identified topological structures must be considered to explain all the emission features associated with the observed flare.
Besides, the HXR sources are regarded as the consequence of the reconnection within or near the border of the flux rope.
\end{abstract}

\section{Introduction}
Flares are believed to be caused by magnetic reconnection that releases part of the energy stored in a non-potential field (e.g., see review by \citealt{2011Shibata} and references therein).
The morphology and evolution of flare ribbons reflect the process of magnetic reconnection.
In the classical two-dimensional (2D) flare model (CSHKP; \citealt{1964Carmichael, 1968Sturrock, 1974Hirayama, 1976Kopp}), the flare ribbons appear as a consequence of heating of the lower dense atmosphere caused by the impact of energetic particles that are accelerated by magnetic reconnection and propagate downward along the field lines.
As magnetic reconnection going on, new flare loops are formed with their altitude moving up and their footpoints separating from each other perpendicularly to the magnetic polarity inversion line.
However, all flares are actually three-dimensional (3D) in nature (e.g., \citealt{2013Janvier,2014Janvier}).
For example, magnetic configurations containing 3D null-points are observed in many events \citep{2007Luoni, 2009Masson, 2012Sun, 2013Sun, 2006Mandrini, 2014Mandrini} in which the flare ribbons, produced by the 3D null-point reconnection, are differently shaped from the 2D scenario with two straight ribbons elongating in the third dimension.

Generally speaking, a null-point configuration is embedded in a multipolar magnetic field and induces a spine-fan structure around it \citep{1996Priest,2008Moreno,2012Sun,2013Sun,2014Mandrini}.
The behavior of the magnetic field close to the null-point can be described by the Jacobian matrix of the field.
The eigenvalues and the eigenvectors of the Jacobian matrix are used to determine the skeleton of the null-point: fan plane and spine.
For a Jacobian matrix having three eigenvectors with three real and distinct eigenvalues \citep{1996Parnell,1997Parnell}\footnote[3]{\url{In the other two cases, the eigenvalues are either complex or repeated. Both of them can also determine the skeleton structure \citep{1996Parnell, 1997Parnell}.}}, the three eigenvalues have different signs, one being positive and the other two being negative or on the contrary.
This is the consequence of the divergence-free condition, which imposes that the sum of these three eigenvalues vanishes.
The two field lines extending from the null-point in the direction of the singly signed eigenvector are defined as the spines, while the field lines extending on a plane and corresponding to the other two eigenvectors with the same sign constitute the fan separatrix, respectively.
The fan surface divides the field into two domains with an inner spine and an outer spine.
The field-line linkage is discontinuous at the fan separatrix, where intense current is prone to be induced \citep{2002Antiochos, 2003Titov, 2003Galsgaard, 2007Pontin, 2012Pontin}.

Magnetic reconnection can also occur at other topological structures such as separators \citep{2010Parnell,b2010Parnell}, bald-patches \citep{1993Titov,1998Aulanier,1999Delannee,2004Pariat}, or more generally, at quasi-separatrix
layers (see \cite{1999Milano,2005Aulanier,2010WilmotSmith,2013Janvier} for MHD simulations, and \cite{2009Lawrence,2012Gekelman} for laboratory experiments).
Quasi-separatrix layers (QSLs) have been successfully associated with both confined and eruptive flares \citep{1997Demoulin,2012Savcheva,2013Guo}.
They are the 3D generalization of separatrices and correspond to volumes of strong gradients of the magnetic connectivity \citep{1996Demoulin, b1996Demoulin, 1997Mandrini, 2002Titov, 2006Demoulin}.
Therefore, the field line connectivities in the two sides of the QSLs are very different.
Usually, the QSLs are narrow layers.
In particular, the thickness of QSLs in a flaring configuration is expected to be of the order of 1 km \citep{2006Demoulin}, which is beyond the resolution limit in both observations and computations.
For viewing purpose, one can select a lower threshold of the squashing factor $Q$, which is a metric of the gradient of the magnetic field line mapping \citep{2002Titov}, that yields a finite thickness of QSLs.
For example, in some cases containing a null-point structure, the spine and fan footpoints are shown to be surrounded by QSLs of a finite thickness \citep{2009Masson, 2012Masson, 2013Sun}.
In observations, flare ribbons always correspond to the intersection of QSLs with the upper photosphere \citep{1991Mandrini, 1994Demoulin}.
Thus, tracing the morphology and evolution of the flare ribbons provides clues to the reconnection process \citep{1988Gorbachev}.

In observations, null-point-related flares would show one or several of the following features.
First, circular ribbons that are associated with 3D null-point would appear at the intersection of the fan with the photoshpere \citep{2006Liu,2009Masson,2012Reid,2012Wang,2013Sun}.
Second, if both the inner and the outer spines of the null are rooted to the photosphere, a short central ribbon and a remote brightening, corresponding to the footprints of the spines, would appear inside and outside the circular ribbon, respectively \citep{2006Liu,2009Masson,2012Reid,2012Wang,2013Sun}.
Two other well-known types of eruptive events could also be observed if a twisted flux rope is present under the fan of the null.
When the flux rope under the fan structure loses its equilibrium, its eruption can generate a blowout jet \citep{2010Moore,2010Sterling,2012Sun,2013Schmieder}.
This would happen when the reconnection in the structure associated with the null-point leads to the disruption of the flux rope \citep{2012Sun,2013Schmieder}.
In such a case, one would observe untwisting plasma motions as the flux rope's disruption results in the launch of non-linear torsional Aflv{\'e}n waves propagating along the newly reconnected field lines \citep{2013Schmieder, 2009Pariat, 2010Pariat, 2015Pariat}.
On the other hand, if the flux rope maintains its integrity, then the null-point reconnection may drive a coronal mass ejection providing that the flux rope successfully erupts \citep{1999Antiochos, 2005Deng, 2008Lynch}.

Recently, the magnetic flux rope has been found likely to appear as a hot channel structure \citep{2012Ding, 2012Cheng, 2013Cheng, 2014Cheng, b2014Cheng}.
The hot channel structure and the prominence correspond to the main body and the lower part of the flux rope, respectively.
Caused by the quick expansion of the flux rope, the upper part of the hot channel seems to separate from the associated prominence.
Meanwhile, both of the hot channel and prominence experience the same morphology transformation \citep{2014Cheng}.
A 3D magnetohydrodynamic simulation of a coronal flux rope by \cite{2012Fan} also supports the above observations.
In the simulation, the 3D morphology of the hot channel is outlined by the isosurface of the temperature at 1.2 MK (Figure 7 in \citealt{2012Fan}).
The hot channel is on the top of the central vertical current layer.
Thus, the hot channel structure is a signature that can be used to trace erupting flux ropes even when there is no filament/prominence, which requires further material condensation \citep{2011Xia, 2012Xia, 2014Xiab, 2014Xiaa}.

In this paper, we study the emission features and, in particular, the magnetic topology of a circular ribbon flare that occurred on 2012 October 23.
We use a nonlinear force-free field extrapolation technique to get the 3D magnetic field over the flaring region and further deduce the 3D squashing factor $Q$ with the state-of-the-art method proposed in \cite{2012Pariat}.
The magnetic field does contain a null-point and a flux rope below the fan structure.
We use the magnetic topology to interpret the multi-wavelength emission features of the flare, especially its flare ribbons and hard X-ray sources. 
This paper is organized as follows.
The observations at different wavelengths are introduced and discussed in Section \ref{Sect.2}.
The magnetic field observations and its topology related to the flare are described in Section \ref{Sect.3}, followed by a discussion and conclusion in Section \ref{Sect.4}.
\section{Multi-wavelength Observations of the Flare} \label{Sect.2}
\subsection{\ion{Ca}{2} H Observations}
On 2012 October 23, an X1.8 class flare occurred in the active region 11598 (S10, E42) that contained two sunspots. 
In \textit{GOES} observations, the flare started at 03:13:00 UT, peaked at 03:17:00 UT, and ended at 03:21:00 UT.
For this event, we study the images of \ion{Ca}{2} H line that come from the Solar Optical Telescope (SOT; \citealt{2008Tsuneta}) on board the \textit{Hinode} satellite.
The data have a spatial scale of $\sim 0.1\arcsec$ per pixel.
Since the coordinate systems of the SOT and other instruments are not consistent with each other, we align the \ion{Ca}{2} H image with a shift of $(25.9\arcsec,20.4\arcsec)$ by calculating the cross-correlation between the SOT \ion{Ca}{2} H image and AIA 1600 \AA \ image.

The central ribbon (R1) and the western part of the quasi-circular ribbon (R2) brightened first at about 03:14:40 UT (Figure \ref{fig:1}).
At the initial time, the two ribbons were very close to each other so that they were visually seen as one ribbon (Figures \ref{fig:1}a--b).
However, after carefully checking the two ribbons in the \ion{Ca}{2} H movie with high temporal resolution (attached to Figure \ref{fig:1}), we find that the two ribbons were not connected, but just separated at a place (marked by the circular symbol in Figure \ref{fig:1}a), with the northern part belonging to the circular ribbon and the southern part to the central ribbon.
The two ribbons, with their footpoints close to the polarity inversion line, suggest that the initial magnetic reconnection occurred at a relatively low level of the atmosphere.
With time going on, the ribbons separated from each other with the reconnection site moving up gradually (Figures \ref{fig:1}b--d and the movie of \ion{Ca}{2} H), just like what is predicted in the CSHKP model.
After the appearance of both the central and quasi-circular ribbons, the remote brightening (R3, Figures \ref{fig:1}c--d) started to brighten, with a time delay of less than 20 s that is the time resolution of the SOT \ion{Ca}{2} H observations.
The SOT \ion{Ca}{2} H images showed three ribbons around the peak time of the flare (Figures \ref{fig:1}c).
The observations displayed some typical features of the circular ribbon flare.
There existed three brightening parts, a central ribbon (R1), a quasi-circular ribbon (R2) curved around the former, and a remote brightening (R3, c.f., \citealt{2009Masson, 2012Wang, 2013Deng, 2013Sun, 2013Liu}).
However, in our event, the remote brightening was quite elongated and seemed to be a third ribbon.

We extract the lightcurves of the \ion{Ca}{2} H ribbons by averaging the intensity over the ribbons and normalizing them by the pre-flare values, with the saturated values being corrected using a linear interpolation.
The selected regions for each ribbon are based on the $20\%$ contour level of the peak value at 03:22:31 UT.
In particular, when the two ribbons are close to each other, we use the magnetic polarities to discriminate them and ensure that the selected regions do not extend out each ribbon.
To estimate the uncertainties of the lightcurves , we select different regions ten times and calculate the standard deviations that are found to be less than $6\%$ of the mean values.
From the lightcurves of the flare ribbons (Figure \ref{fig:2}), the integrated emissions of the three ribbons display a time sequence in the rise phase: the first is the central ribbon, followed by the quasi-circular ribbon, and the last is the remote ribbon.
Though the emission of the remote ribbon rises last, it gets to peak earlier than the other two ribbons.
Such a brightening sequence is due to the process of the flare reconnection as will be discussed in Section \ref{Sect.4}.

\subsection{Extreme Ultraviolet Observations}\label{Sect.2.2}
The 304 \AA , 171 \AA \ and 94 \AA \ images from the Atmospheric Imaging Assembly (AIA; \citealt{2012Lemen}) on board the \textit{Solar Dynamics Observatory} (\textit{SDO}) give a different view of this flare from the \ion{Ca}{2} H images.
The images of the AIA 304 \AA \ passband display emission features from a low characteristic temperature of $5.0\times 10^4$ K, which includes the \ion{He}{2} line formed in the chromosphere and the transition region.
The \ion{Fe}{9} line emission is contained in the AIA 171 \AA \ passband that originates from the quiet corona and the upper transition region, with a characteristic temperature of $6.3\times 10^{5}$ K.
The AIA 94 \AA \ passband, containing the \ion{Fe}{18} line emission, originates from the flaring region corona, with a characteristic temperature of $6.3\times10^{6}$ K.
All the observations in these passbands (Figure \ref{fig:3}) with characteristic temperatures ranging over 2 orders of magnitude show the evolution of the flare in the higher atmosphere than the \ion{Ca}{2} H images.

One can find that at the initial time, a bright arcade-like structure (ALS) lay low with its two footpoints anchored in the solar surface (Figures \ref{fig:3}a, \ref{fig:3}e, and \ref{fig:3}i).
From the observations of AIA 94 \AA , three other structures can be found there: highly sheared field lines just above the ALS, a dome-like structure over the ALS and the sheared field lines, a loop-like structure that connected the dome-like structure and the western part of the solar surface.
After a short time, at about 03:15:00 UT, this bright ALS rose suddenly and got much brighter (Figures \ref{fig:3}b, \ref{fig:3}f, \ref{fig:3}j, and the movie attached to Figure \ref{fig:3}).
Meanwhile, the sheared field lines were squeezed by the rising ALS, and their northern ends also got brighter. 
Moreover, the rising ALS seems to have a little twist, as shown especially in Figure \ref{fig:3}f and the movie (the AIA 171 \AA \ emission) attached to Figure \ref{fig:3}, implying that it may be a flux rope.
With time going on, at about 03:15:25 UT, the rising ALS got higher and it then pushed the dome-like structure over it, as revealed in the movie (the AIA 94 \AA \ emission) attached to Figure \ref{fig:3}.
Simultaneously with this process, the circular ribbon appeared, which, however, can only be seen in the high temperature emissions initially (Figure \ref{fig:3}k).
Such an evolution process suggests that the squeezing of the dome-like structure by the ALS may cause a magnetic reconnection between them, and then the accelerated electrons propagate along the field lines of the dome-like structure to heat the lower atmosphere that forms the circular ribbon.
As the ALS rose further up, it became less bright, while the remote and the circular ribbons can be seen in the images of all the three wavebands (Figures \ref{fig:3}d, \ref{fig:3}h, and \ref{fig:3}l).
It is worth noting that during the process of the ALS rising, the bright loop-like structure was dragged to a higher place (orange dotted line in Figures \ref{fig:3}i--l).
Then, the remote ribbon appeared just at the western end of this loop-like structure at about 03:15:50 UT (Figure \ref{fig:3}l).
This implies that energetic electrons are transported through the field lines in this large loop-like structure to the lower atmosphere and the remote ribbon is caused by the collision of the energetic electrons with the chromospheric plasma.
In addition, with the ALS rising, its southern footpoint experienced a slight movement but was still anchored inside the circular ribbon.
There are two possibilities for this motion: one is the lateral expansion of the flux rope during its eruption; the other is magnetic reconnection of the flux rope with the ambient field that causes an apparent motion of the footpoints.
The evolution of the flare in AIA 94 \AA \ is also shown by the movie attached to Figure \ref{fig:3}.
Comparing the highly sheared loops in AIA 94 \AA \ at the initial time and the flare loops in \ion{Ca}{2} H at a later time, one can find that the shear becomes weaker after the flare (Figure \ref{fig:1} and the \ion{Ca}{2} H movie attached to it).
Such a  weakening of the shear angle in a flare process has been studied by many authors \citep{2003Asai,2010LiuR,2011InoueS,2011WarrenHP,2012SavageSL,2012Aulanier}.

The bright ALS is identified as a hot channel that likely corresponds to a flux rope \citep{2012Fan,2014Cheng}.
More evidence for the existence of flux rope can be found in Section \ref{Sect.3}.
The initial rising of the ALS might be a consequence of tether-cutting type reconnection \citep{2001Moore}, breakout reconnection \citep{1999Antiochos}, torus instability \citep{2006Kliem}, kink instability \citep{2005Torok,2010Guo}, flux emergence \citep{2000Chen}, or magnetic flux cancellation \citep{1993WangJX,2013Burtseva}.
This dynamic process drove the flare to proceed quickly; the 304 \AA , 171 \AA , and 94 \AA \ emissions reached to the peak in about 2 minutes and caused the saturation of the detectors.
Moreover, as shown in the AIA 94 \AA \ movie, with the ALS rising, it became fainter and fainter and finally hard to be seen.
Then, it is hard to judge what happened to the flux rope in the later stage.
It might undergo a full eruption or only part of it erupted.
Nevertheless, no coronal mass ejection was observed during the flare by the Large Angle and Spectrometric Coronagraph C2 on the Solar and Heliospheric Observatory.

\subsection{Hard X-ray Emission Features}\label{Sect.2.3}
Hard X-ray observations of the flare were recorded by the \textit{Ramaty High Energy Solar Spectroscopic Imager} (\textit{RHESSI}; \citealt{2002Lin}) and started at 03:14:36 UT.
We reconstruct the hard X-ray sources by the Clean algorithm \citep{1974Hogbom,2002Hurford} with 1F, 3F, 4F, 5F, 6F, 7F, and 8F detectors for two energy bands of 12--25 keV and 50--100 keV, which are superimposed on the \ion{Ca}{2} H, 304 \AA , 171 \AA , and 94 \AA \ images in Figure \ref{fig:1} and Figure \ref{fig:3}.
Two HXR sources can be seen in the 12--25 keV energy band in some of these images.
The fluxes of the two HXR sources X1 and X2 are calculated by summing all the pixels in the white boxes (Figure \ref{fig:2}a) and then normalized to their peak values (Figure \ref{fig:2}b).

Contrary to the \ion{Ca}{2} H emission that was ribbon-like, the HXR sources were localized at some particular sites, as usual \citep{2002Asai,2007Temmer,2007Miklenic,2012Guo}. 
The source X1 was located near the northern end of the parallel part of the central and quasi-circular \ion{Ca}{2} H ribbons, while source X2 was in the southern part of the central ribbon (Figure \ref{fig:1}c).
The HXR source in the 50--100 keV energy band overlay the parallel part of the central and quasi-circular ribbons.
What is more interesting is that this HXR source showed an obvious southward movement along the elongated parallel ribbons (Figures \ref{fig:1}b--d), which is somewhat different from the scenario found by \citet{2012Reid}, in which the main HXR source, associated with the inner spine and the fan surface, keeps stationary for about 60 s, although the secondary source is actually not stationary.
Meanwhile, the two HXR sources in 12--25 keV also experienced a movement along the central ribbon (Figure \ref{fig:2}a).
The source X2 moved about $10\arcsec$ toward south, while the motion of source X1 was less than $5\arcsec$ toward south.
Note that the cases of both the HXR sources moving parallel to the polarity inversion line are not rare, as revealed in a statistic study by \citet{2005Bogachev}.
The authors explained such cases as being caused by the reconnection successively occurring along the separator.
In those cases, the distance of the two HXR sources do not change obviously with time.
However, the case here is somewhat different in that the distance of the two HXR sources changed obviously with time, with the northern one being nearly static but the southern one moving to south.

In order to make a proper interpretation for the two HXR sources, we further compare the HXR images with the EUV emissions at 304 \AA , 171 \AA , and 94 \AA \ (Figure \ref{fig:3}).
The result shows that two HXR sources correspond to the two footpoints of the rising flux rope, as judged from both the hot channel structure (outlined by the dotted line in Figure \ref{fig:3}) and the magnetic field extrapolation (shown in Section \ref{Sect.3.2}).
The northern footpoint of the flux rope moved little and had a good correlation with the northern HXR source.
The southern HXR source moved toward south and so did the southern footpoint of the flux rope in the EUV images, although their locations were not exactly cospatial.
Such a spatial relationship and similar motion patterns suggest that the HXR sources may be caused by the magnetic reconnection within or near the border of the flux rope.
Besides, the fluxes of the two HXR sources were not peaked simultaneously.
The peak of the source X2 was delayed by about 1 minute with respect to the peak of source X1 (Figure \ref{fig:2}b).

\section{Magnetic Field Topology of the Flare}\label{Sect.3}
\subsection{Magnetic Field Observation and Extrapolation}\label{Sect.3.1}
The photospheric vector magnetic field data of the active region were observed by the Helioseismic and Magnetic Imager (HMI; \citealt{2012Schou, 2012Scherrer}) on board \textit{SDO}.
The data have a pixel spacing of $\sim 0.5\arcsec$.
As shown in Figure \ref{fig:4}, the magnetic flux distribution of the active region is, to a first order, equivalent to typical quadrupolar magnetic fields, with the four poles defined as P1, P2, N1, and N2, respectively.
The positive pole P1 was located on the western side, far from the other three poles on the eastern side (P2 lies in between N1 and N2).

We extrapolate the coronal magnetic field by the nonlinear force-free field model with the optimization method \citep{2000Wheatland, 2004Wiegelmann} using the magnetogram observed at 02:59:58 UT before the flare (Figure \ref{fig:4}a).
First, the 180\arcdeg \ ambiguity of the transverse components of the magnetic field is removed with the minimum energy method \citep{1994Metcalf, 2006Metcalf, 2009Leka}.
Since the active region is not located close to the disk center, the projection effect should be corrected with the method proposed by \citet{1990Gary}.
As a consequence of removing the projection effect, the geometry of the field of view is changed (Figure \ref{fig:4}b).
We recut the edges to get a rectangle boundary for the extrapolation.
The extrapolated area is resolved by $ 200\times178 $ grid points with $\delta x = \delta y \approx 1\arcsec$.
Moreover, considering that the force-free and torque-free conditions are usually not satisfied for the observed photospheric magnetic field, a preprocessing is applied to remove the net force and torque on the boundary \citep{2006Wiegelmann}.
The vector magnetic field at the bottom boundary after the above preprocessing is shown in Figure \ref{fig:4}c.

Selected field lines from the resulting extrapolation are shown in Figure \ref{fig:5}.
In order to check the force-free state of the extrapolated result, we choose the force-freeness metric used by \cite{2000Wheatland}, the sine of the angle between the current density and the magnetic field averaged by current magnitude:
\begin{equation}
\langle\mathbf{CW} \mathbf{sin}\theta\rangle = \frac{\sum_{i}J_i \mathbf{sin}\theta_i}{\sum_{i}J_i},
\end{equation}
where
\begin{equation}
\mathbf{sin}\theta_i = \frac{|\mathbf{J}_i\times\mathbf{B}_i|}{J_i B_i}.
\end{equation}
The metrics of the force-freeness of the extrapolation for the whole 3D computational domain and the box containing the flux rope  (shown in Figure \ref{fig:5}c) are $\sim 0.41$ and $\sim 0.19$, corresponding to a current weighted angle of $\sim 24^{\circ}$ and $\sim 11^{\circ}$, respectively.
The departure of the result from the force-freeness might be caused by the boundary condition, which does not satisfy the force-free condition strictly, even after the preprocessing.
Adoption of a planar geometry of the bottom boundary may also have some influence, since the computational domain is quite large and the curvature effect cannot be fully neglected.
Nevertheless, the current weighted angles in our extrapolation are similar to previous studies (e.g., \citealt{2008Schrijver,2010Guo,2012Sunb}).

\subsection{Topological Analysis} \label{Sect.3.2}
Using the extrapolated 3D coronal magnetic field,
we solve the equation $B_{i}(x, y, z) = 0$ (where $i = x, y, z$) with the modified Powell hybrid method\footnote[4]{\url{http://www.lesia.obspm.fr/fromage/}}.
The magnetic field in the vicinity of the null is described as the linear term $\mathbf{B}=\mathbf{M}\cdot\mathbf{r}$, where $\mathbf{M}$ is the Jacobian matrix for the magnetic field ($M_{ij}=\partial B_{i}/\partial x_{j}$), and $\mathbf{r}$ is the vector of the position.
Only one null-point is found in the domain, which is marked by a white star in Figure \ref{fig:5}. 
The skeleton structure of the null-point, to say, the field lines integrated along the eigenvectors of the Jacobian matrix from the null-point, is shown as the red field lines in Figure \ref{fig:5}.
Next, we calculate the squashing factor $Q$ with the method proposed by \citet{2012Pariat}.
This parameter is a geometric measure of the deformation of the elementary flux tube cross section and gives the most important information on the magnetic field connectivity gradient.
The region with high but not infinite $Q$ values is usually a 3D volume with finite thickness.
By comparison, the separatrices belong to a limit case with the squashing factor $Q$ going to infinity and the thickness being infinitesimal.
The QSL contains both the regions with drastic ($2\ll Q\not= \infty$) and discontinuous ($Q = \infty$) connectivities of magnetic field lines.
We also calculated the QSLs derived from a potential field extrapolation to emphasize some of the identified QSL structures and their interconnections (Figure \ref{fig:5}f and Figure \ref{fig:6}).

The topology of the magnetic field reveals three structures as can be seen in Figure \ref{fig:5} and Figure \ref{fig:6}.
First, a large-scale QSL, induced by the large quadrupolar-like magnetic fields, appears like the cases studied by \cite{1996Demoulin} and \cite{2005Aulanier}, in which the footprints of the QSL have two thin strips over two magnetic polarities, respectively.
Second, we find a single 3D null-point embedded within the large-scale QSL.
Third, a flux rope is present under the fan structure.
The footprints of the large-scale QSL are two C-shaped and elongated regions (C1 and C2), corresponding to the footpoints of the yellow field lines (Figures \ref{fig:5}a--b) and also appearing in the squashing factor map (Figures \ref{fig:5}e--f and Figure \ref{fig:6}).
The single null-point is located above a place to the south of the positive pole P2, the inner spine is rooted in the south of P2, and the outer spine is rooted in a parasitic positive pole to the south of the negative pole N1.
The fan structure divides the volume into inner and outer domains, and the intersection of the fan with the lower boundary forms a quasi-circular morphology (fan QSL, Figures \ref{fig:5}e--f) that is associated with the quasi-circular ribbon in AIA 1600 \AA .
For a better comparison, the AIA 1600 \AA \ image has been processed by removing the projection effect (Figures \ref{fig:5}b and \ref{fig:6}) with the method proposed by \citet{1990Gary}.
The footprint of the fan structure divides the C1 structure into two parts, an inner one and an outer one (Figures \ref{fig:5}e--f), which correspond to the inner ribbon and the remote ribbon in 1600 \AA, respectively.
The eastern C-shaped region (C2) in the $Q$ map (Figures \ref{fig:5}e--f and Figure 6) corresponds to the weak ribbon appearing in AIA 1600 \AA \ (Figure \ref{fig:5}b and Figure \ref{fig:6}) but hardly seen in \ion{Ca}{2} H observations (Figure \ref{fig:1}).
The flux rope lies along the western part of the P2-N1 polarity inversion line.
Moreover, the two hard X-ray sources in 12--25 keV are located beneath the flux rope (Figure \ref{fig:5}c).

To view more clearly the QSL topology, we plot in Figure 7 the distribution of the squashing factor $Q$ in the $x-y$  plane for $z=0$ and in the $x-z$ plane for different values of $y$.
It is seen that viewed in the $x-z$ plane, the squashing factor $Q$ shows a fan structure (Figures \ref{fig:7}b--c) and also a second dome-like structure above the fan of the null-point (denoted as L2 in Figures \ref{fig:7}h).
Compared with the 3D view of the $Q$ halo (Figure \ref{fig:6} and the movie attached to Figure \ref{fig:6}), L2 corresponds to the cut in the $x-z$ plane of eastern structure (S2) of the QSL induced by the large-scale quadrupolar-like magnetic field.
L2 is partially coincident with the fan as shown in Figure \ref{fig:7}b--d.
On the other hand, a vertical structure in the $x-z$ plane (denoted as L1 in Figure \ref{fig:7}h) corresponds to the western structure (S1) of the QSL.
And the inner domain of the fan consists of two parts, the eastern part (EP) and the western part (WP).
The flux rope lies in the latter.
Such a structure of double parts is similar to the `fish-bone-like' structure induced by a magnetic null line \citep{2014Wang}; however, we do not find a null-line in the magnetic structure of this event.
On the other hand, we can see that the single null-point is embedded in the western structure of the large-scale QSL (S1 in Figure \ref{fig:6} and L1 in Figure \ref{fig:7}i).
Thus, the basic magnetic topology of this active region is a null-point with its related skeleton embedded in a large-scale QSL and a flux rope lying below the null-point-related fan structure.

Since there is a possibility that the flux rope finally erupts, we check the magnetic structure after the flare with NLFFF extrapolation and find that the twisted flux rope seems to be replaced by a sheared magnetic field (Figure \ref{fig:8}a).
The 3D null-point also disappears.
However, the double C-shaped footprint of the large-scale QSL still remains (C1 and C2), as shown in both the $Q$ maps of the NLFFF and potential field (Figures \ref{fig:8}b--c).

\section{Discussion and Conclusion} \label{Sect.4}
The flare studied in this paper was associated with a complex magnetic topology comprising a flux rope, a 3D null-point, and a large-scale QSL. The flux rope was initially located below the fan of the null-point while the null-point was itself embedded in the large-scale QSL.
The emission features of the flare in \ion{Ca}{2} H line, 304 \AA , 171 \AA , and 94 \AA , and HXR are also presented in order to study its occurrence and evolution, with regard to the aforementioned complex magnetic topology.
Here, we highlight three points:
first, we speculate that the HXR sources are caused by the reconnection within or near the border of the flux rope; 
secondly, the evolution of the inner and circular ribbons is associated successively with the footprints of the flux rope QSL and the spine-fan QSL;
thirdly, a dome-like and a loop-like structures in AIA 94 \AA \ are observed to correspond to the 3D large-scale QSL halo.

We identify the ALS (Figure \ref{fig:3}) as the flux rope by comparing it with the extrapolated magnetic field (Figure \ref{fig:5}).
Further evidence for the flux rope identification is the hot channel structure in the 304 \AA , 171 \AA , and 94 \AA \ images (Figure \ref{fig:3}).
The hot channel structure can be a part of the magnetic flux rope \citep{2012Fan, 2012Ding, 2012Cheng, 2013Cheng, 2014Cheng, b2014Cheng}.

In order to give a proper interpretation for the HXR emission, we make a comparison of the HXR sources with the 304 \AA , 171 \AA , and 94 \AA \ images.
The result shows that the two HXR sources were located near the footpoints of the flux rope.
Further analysis of the extrapolated 3D magnetic field (Figures \ref{fig:5}c) confirms the spatial relationship between the HXR sources and the flux rope footpoints.
In particular, with the flux rope rising, its footpoints experienced a slight movement (Figure \ref{fig:3}) and the HXR sources showed a southward movement along the polarity inversion line (Figure \ref{fig:2}).
The possible origins of this slight movement are discussed in Section \ref{Sect.2.3}.
Regardless of these motions, the two HXR sources were still located near the footpoints of the flux rope.
This is to say, although they are not exactly cospatial, their motion patterns are similar.
Therefore, the HXR sources are considered to be related to the flux rope footpoints.

A number of previous studies have been done to investigate the magnetic reconnection of a 3D null-point, for example, ideal kinematic models \citep{1996Klapper,1997Bulanov,2003Mellor,2011Pontin}, torsional spine and fan reconnection model that occurs due to a rotational disturbance at the fan plane or around the spine \citep{1996Rickard,2007Pontin,2011Pontin}, and the most common spine-fan reconnection when a shear disturbance of either the spine or the fan occurs \citep{2007aPontin,2011Pontin}.
In the above models, the magnetic null point collapses and generates electric current locally.
The energetic electrons are likely to flow from the null-point along the inner/outer spines and the eigenvector with the largest absolute eigenvalue in the fan of the magnetic field.
However, magnetic reconnection does not solely occur at the null-point, but also possibly in extended places with large $Q$ values, like slipping or slip-running reconnection \citep{2006Aulanier, 2009Masson, 2012Reid, 2013Janvier}.
Comparing the skeleton structure of the null-point and the QSL structure with the \ion{Ca}{2} H emission for the present event, we find that both the footpoints of the inner and the outer spines do not correspond to the major bright region; they lie even far from, for example, the 1600 \AA \ emission (Figure \ref{fig:6}) and, in particular, the HXR emission (Figure \ref{fig:5}c).
Figure \ref{fig:6} shows that the inner spine is rooted in the southern end of the central ribbon.
And the outer spine is rooted in the eastern end of the extremely elongated remote ribbon.
Nevertheless, the QSLs correspond to the ribbons well.

In addition, we find that not only the ribbons in AIA 1600 \AA \ correspond to the intersections of the QSLs with the lower atmosphere, but also those in AIA 94 \AA \ correspond to the QSL halo.
By comparing the 3D QSL configuration (Figure \ref{fig:6}) with the AIA 94 \AA \ emission (Figures \ref{fig:3}i--l), we find that the field lines embedded in the structure S2 correspond to the dome-like struture and those embedded in the structure S1 correspond to the loop-like structure that connects the dome-like structure and the remote ribbon.
Physically, the QSLs are places where intense currents may appear \citep{2003Titov, 2003Galsgaard} and thus dissipation can take place \citep{2005Aulanier,2010WilmotSmith,2013Janvier}.
Therefore, the plasmas in the QSLs could be heated by the dissipation and then appear as a bright structure that can delineate the QSL topology.
This can explain the observational findings in our event.
This event is among the very few revealing both a dome-like structure and a loop-like structure in the EUV emission, which is consistent with a QSL configuration.

\begin{figure}
\centering
\includegraphics[width=0.84\textwidth]{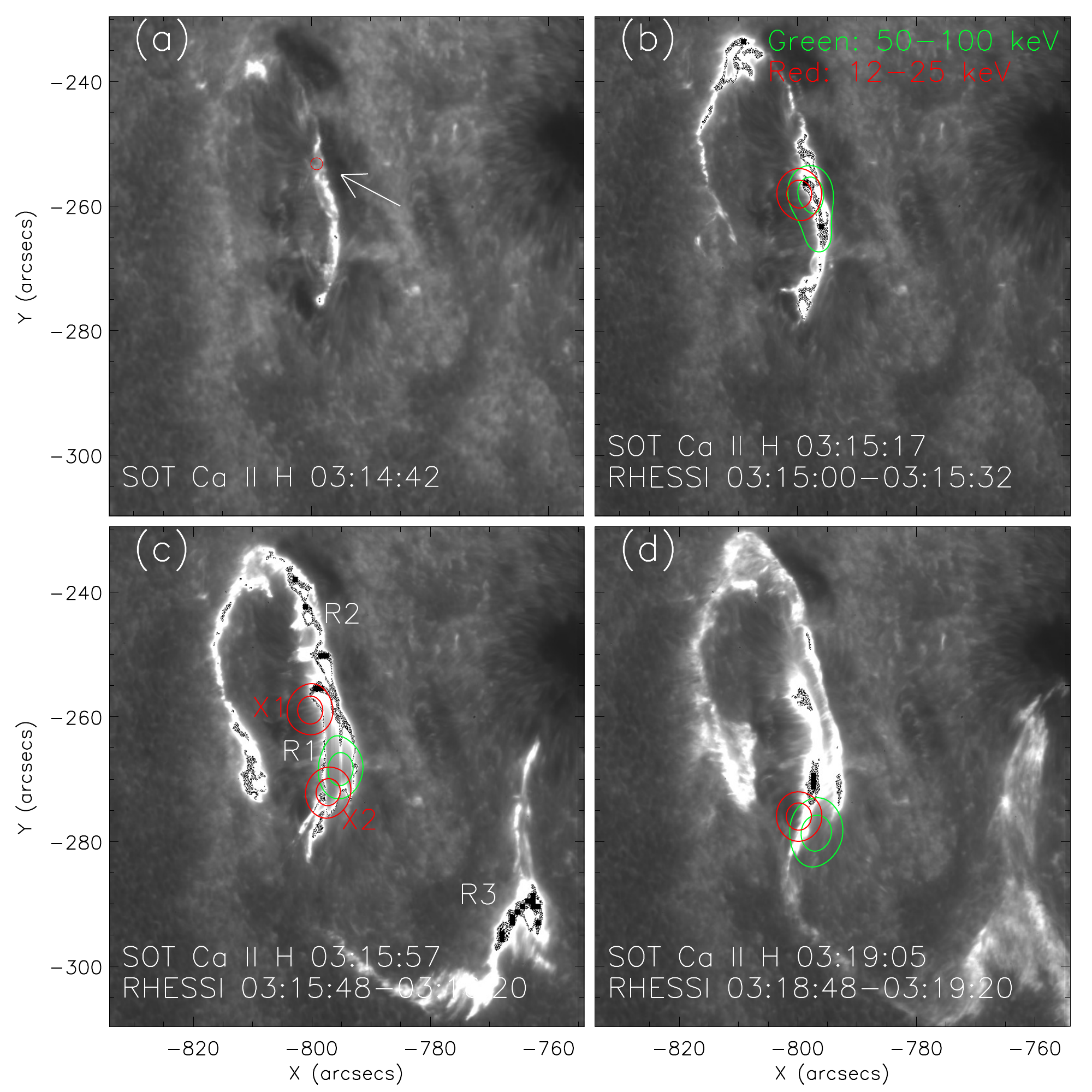} 
\caption{(a)--(d) SOT/Hinode \ion{Ca}{2} H images showing three flare ribbons: a central ribbon (R1), a quasi-circular ribbon (R2), and a remote ribbon (R3). The red circular symbol and the arrow in panel (a) indicate the place where the two ribbons separate from each other. HXR sources in the 12--25 keV and 50--100 keV energy bands at the closest time to the \ion{Ca}{2} H images are overplotted with contour levels of 50\% and 80\% of the peak flux. A \ion{Ca}{2} H movie is attached to this figure.}
\label{fig:1}
\end{figure}

\begin{figure}
\centering
\includegraphics[width=1\textwidth]{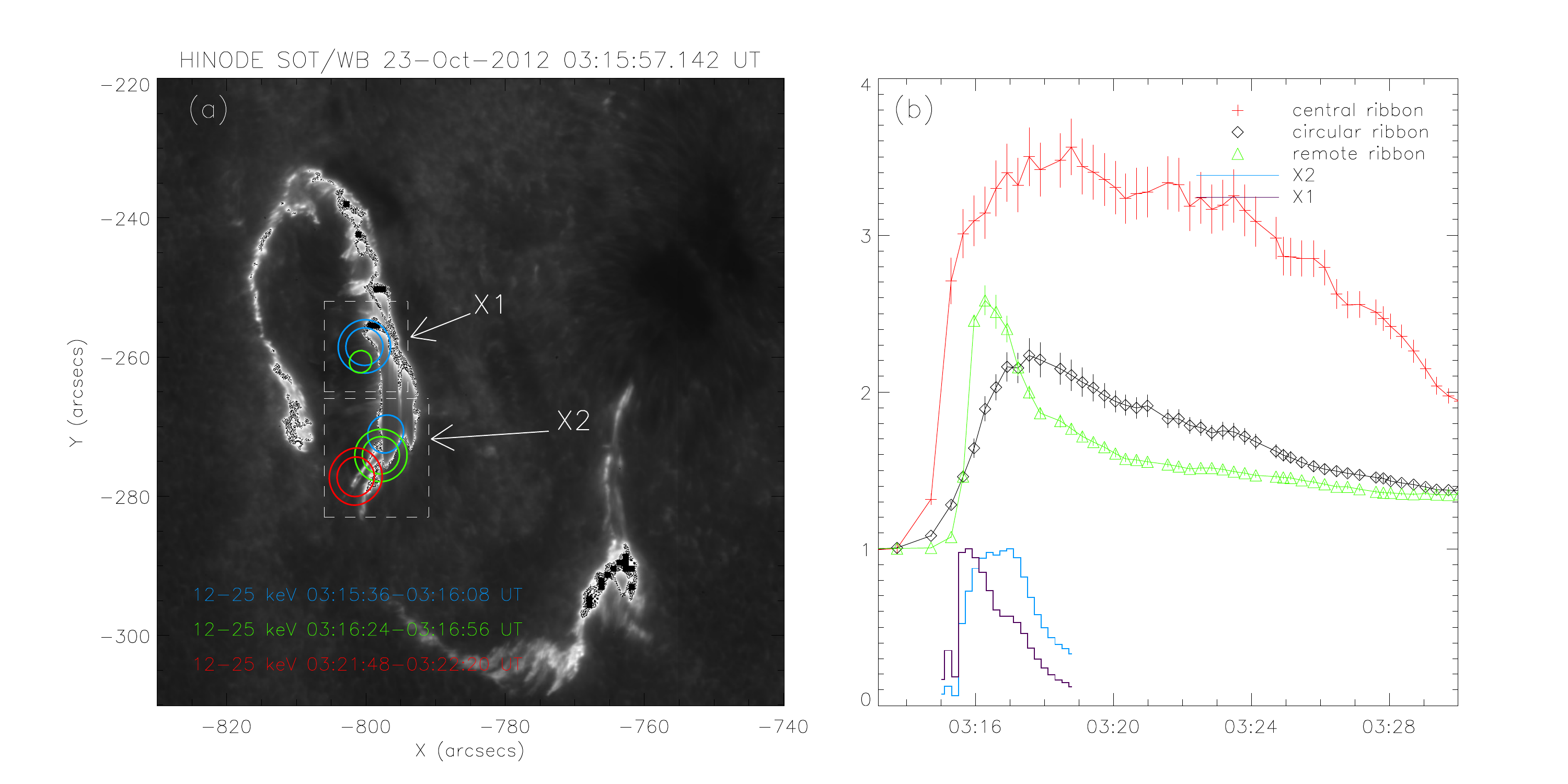}
\caption{ (a) SOT/Hinode \ion{Ca}{2} H image at 03:15:57 UT on 23 October 2012 overlaid with the HXR sources at different times in the energy band of 12--25 keV. The solid lines indicate the contour levels of 50\% and 80\% of the HXR peak flux. Blue, green, and red colors trace the time sequence of 03:15:36 to 03:16:08 UT, 03:16:24 to 03:16:56 UT, and 03:21:48 to 03:22:20 UT, respectively. The two white boxes delineate the regions over which the HXR light curve are extracted. (b) \ion{Ca}{2} H lightcurves of the three flare ribbons and the HXR fluxes of the two sources in the 12--25 keV energy band. The errorbars stand for the standard deviation of the measured fluxes.}
\label{fig:2}
\end{figure}

\begin{figure}
\centering
\includegraphics[width=0.8\textwidth]{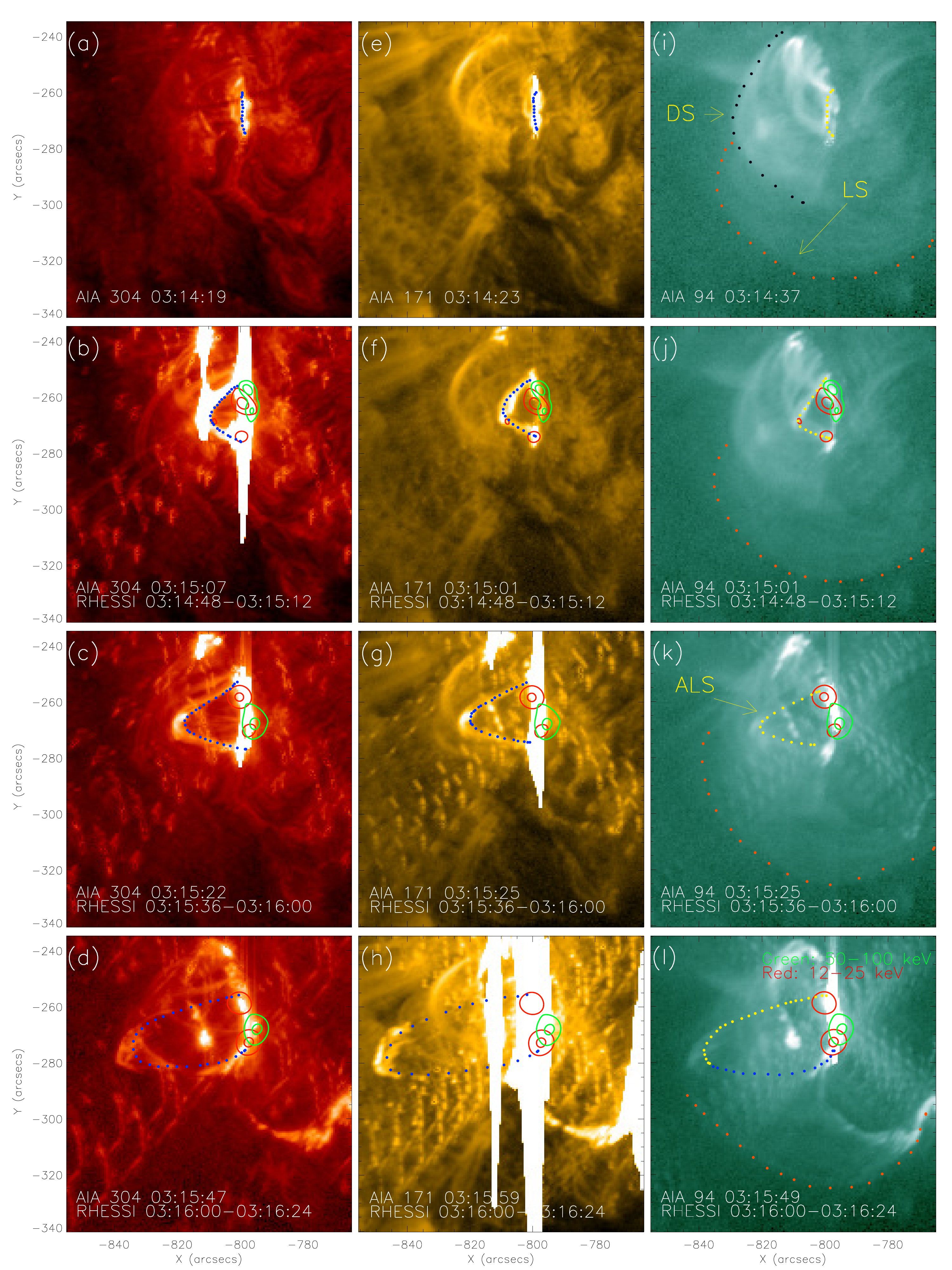}
\caption{ (a)--(d) AIA 304 \AA \ images, (e)--(h) AIA 171 \AA \ images, and (i)--(l) AIA 94 \AA \ images showing the evolution of the flare. The blue and yellow dotted curves in (a)--(l) outline the rising arcade-like structure. In panel (l), the yellow curve outlines the visible part in 94 \AA , and the blue one is plotted by referring to the structure in 304 \AA \ in panel (d). The orange dotted lines in panels (i)--(l) outline the loop-like structure (labelled as LS). The black dotted line in panel (i) outlines the dome-like structure (labelled as DS). The arcade-like structure is labelled as ALS in panel (k). HXR sources in the 12--25 keV and 50--100 keV energy bands are overplotted with contour levels of 50\% and 90\% of the peak flux. A movie of 94 \AA \ emission is attached to this figure. }
\label{fig:3}
\end{figure}

\begin{figure}
\centering
\includegraphics[width=1.0\textwidth]{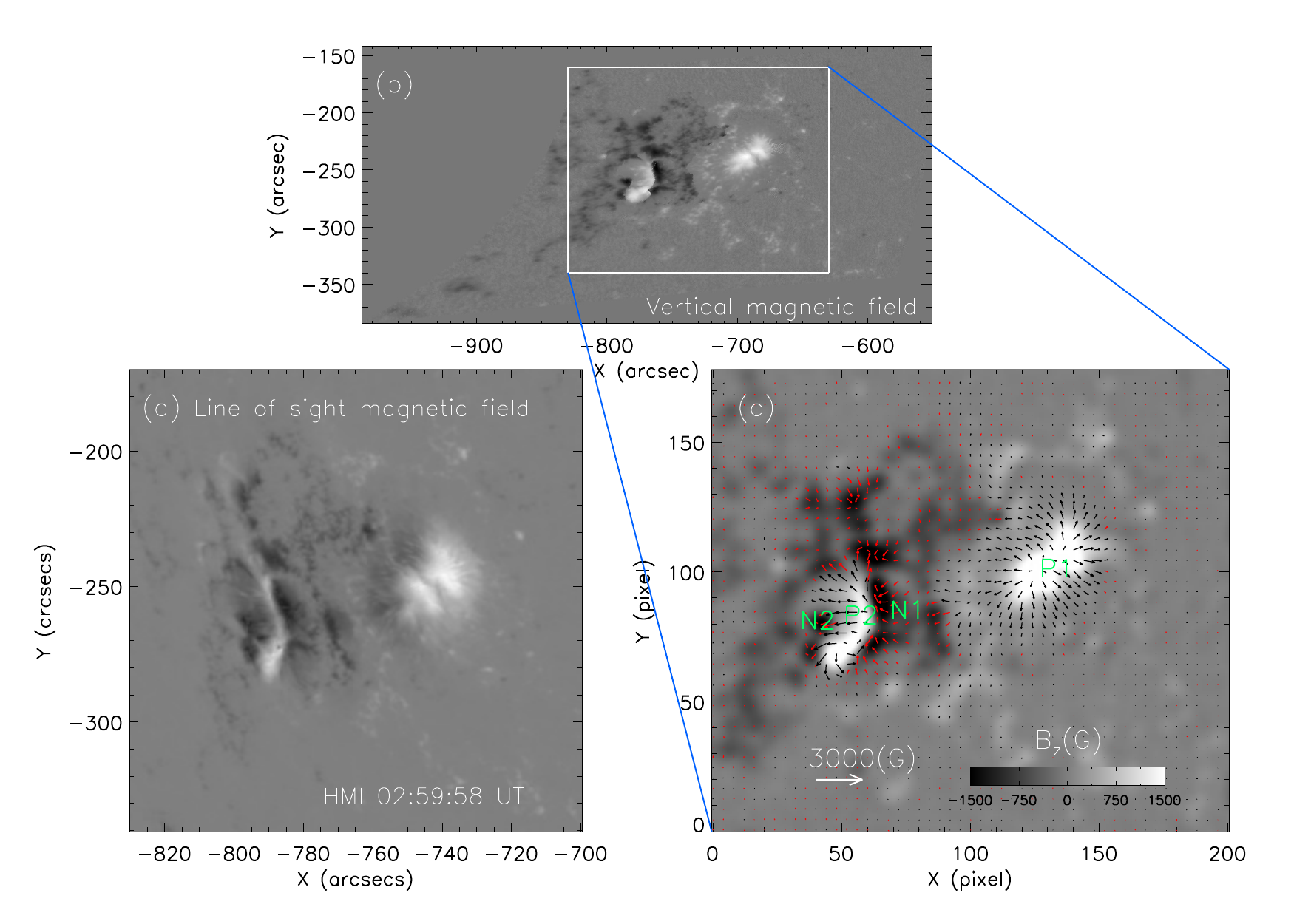}
\caption{(a) Photospheric line-of-sight magnetic field of the active region 11598 at 02:59:58 UT observed by HMI/\textit{SDO} before any preprocessing. (b) The vertical magnetic field after removal of the 180\arcdeg \ ambiguity in the transverse components of the magnetic field and correction for the projection effect. (c) The vector field at the bottom boundary after preprocessing used for the extrapolation. The gray-scale image stands for the vertical component of the magnetic field, while the arrows in panel (c) indicate the magnitude and direction of the horizontal component. The two positive poles and two negative poles are defined as P1, P2, N1, and N2, respectively.}
\label{fig:4}
\end{figure}

\begin{figure}
\centering
\includegraphics[width=0.9\textwidth]{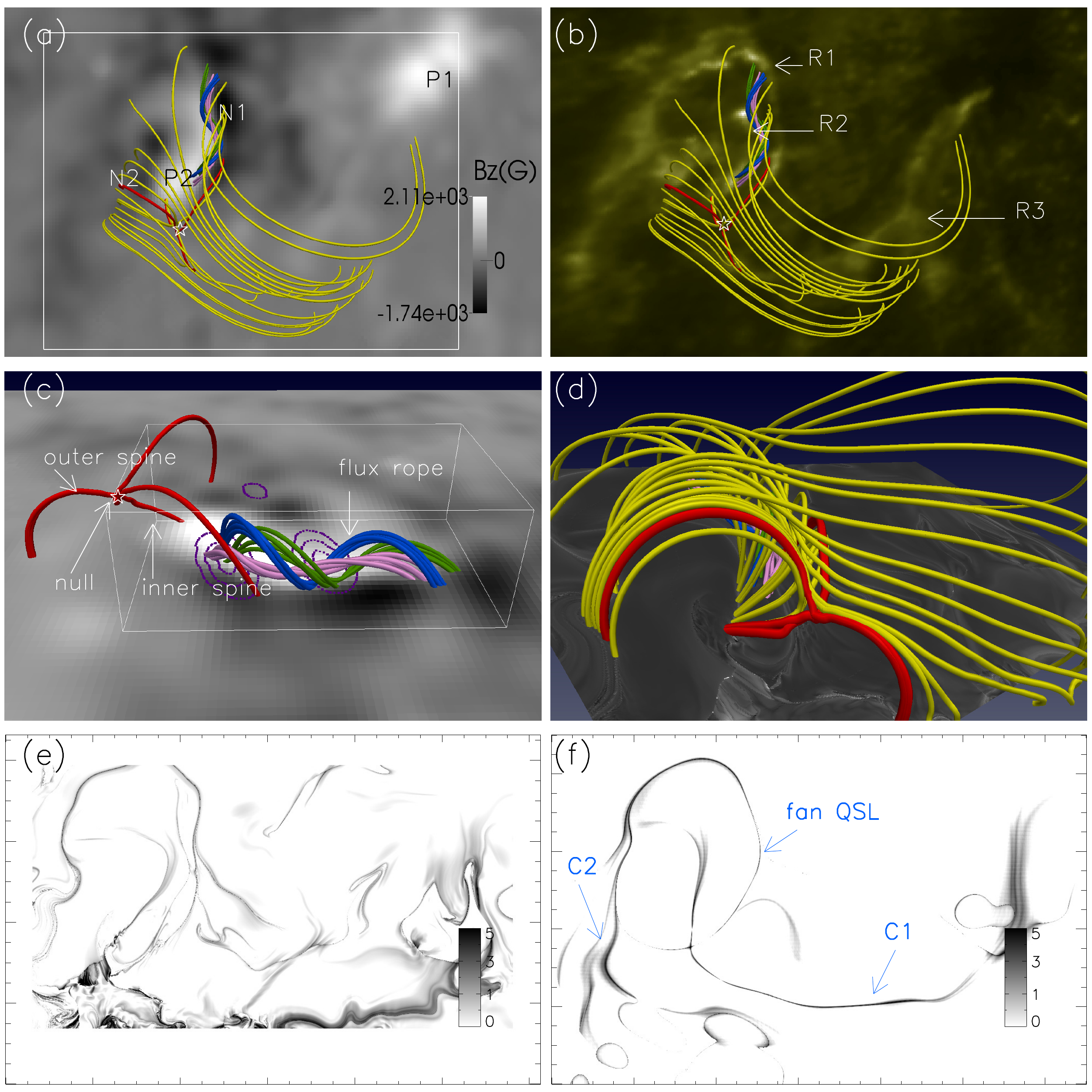}
\caption{(a) Magnetic field lines in the computation box, which depicts the null-point (red), the QSLs  (yellow), and the flux rope (pink-green-blue). The background shows the photospheric vertical magnetic field. The white box shows the computational domain of the squashing factor $Q$. (b) The same as panel (a) with the bottom boundary being replaced with the AIA 1600 \AA \ image at 03:34:40 UT. (c) A side view of panel (a) without the yellow magnetic field lines. HXR sources in the 12--25 keV band, with the time interval of 03:16:00--03:16:30 UT and contour levels of 50\%, 70\%, and 90\% of the peak flux, are overlaid on the photospheric vertical magnetic field. The white box in panel (c) is used to check the force-freeness metric of the flux rope. (d) A side view of panel (a) with the bottom boundary being replaced with the $Q$ map. (e)--(f) Maps of the squashing factor $Q$ on the bottom boundary from NLFFF field and potential field, respectively. The two C-shaped footprints of the large-scale QSL are labelled as C1 and C2, respectively.}
\label{fig:5}
\end{figure}

\begin{figure}
\centering
\includegraphics[width=1\textwidth]{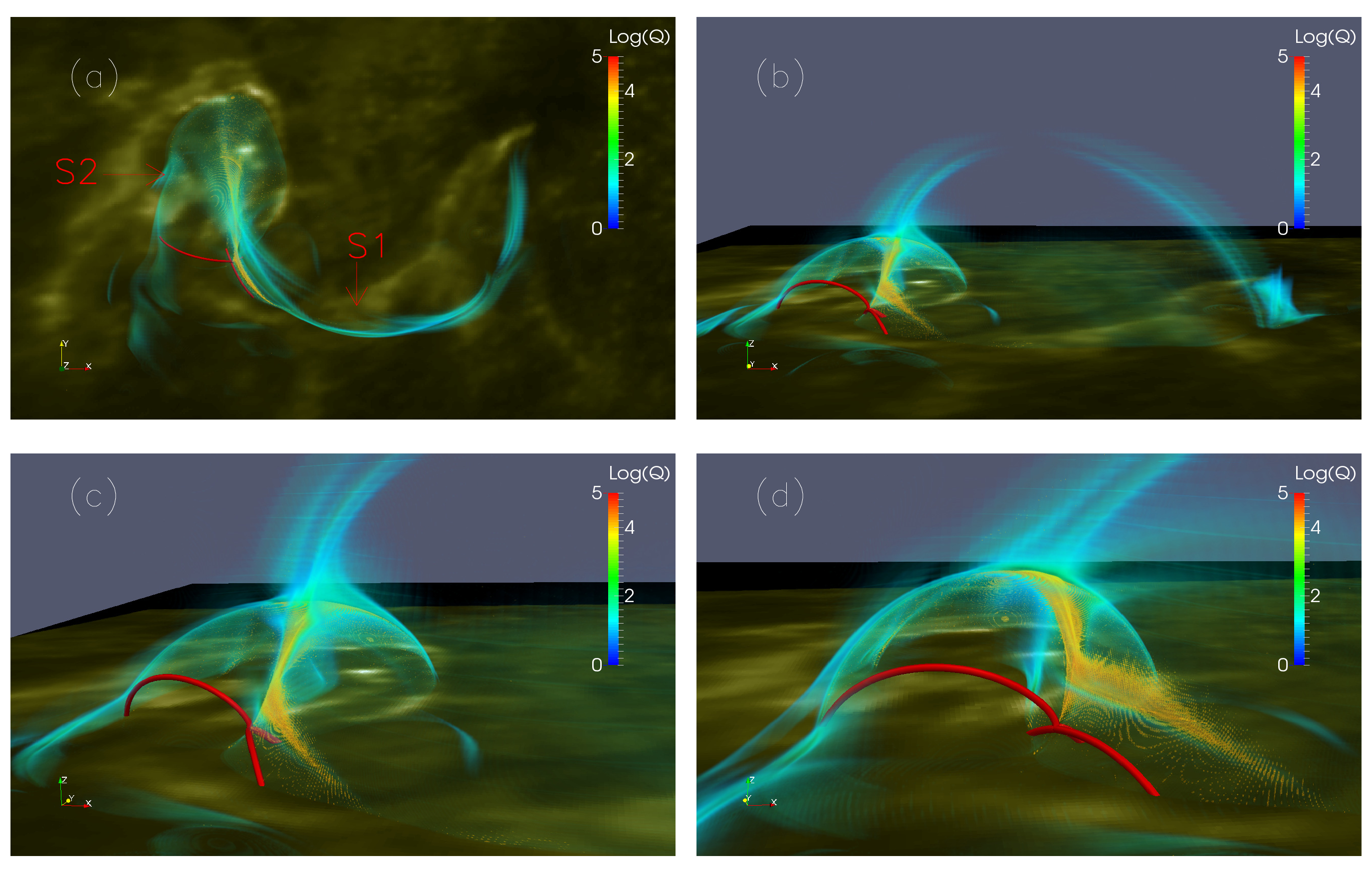}
\caption{(a)--(d) 3D Views of QSL calculated from the potential field. The image on the bottom boundary shows the AIA 1600 \AA \ emission at 03:34:40 UT. The red lines represent the skeleton structure of the null-point. The western and the eastern structures of the large-scale QSL that correspond to the loop-like and dome-like structures in AIA 94 \AA \ are labeled as S1 and S2, respectively. A movie showing the 3D QSL is attached to this figure.}
\label{fig:6}
\end{figure}

\begin{figure}
\centering
\includegraphics[width=1\textwidth]{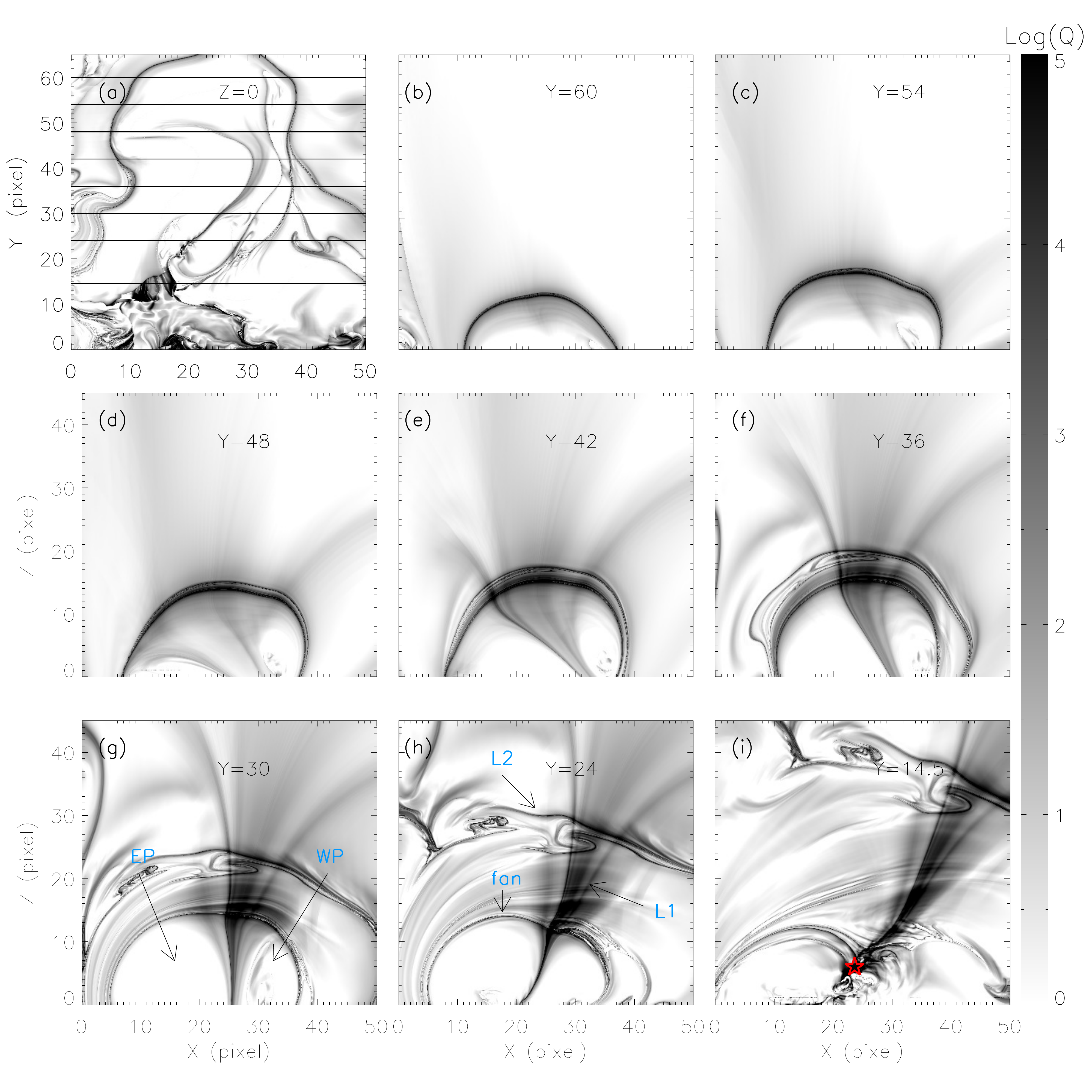}
\caption{(a) Map of the squashing factor $Q$ in the \textit{x--y} plane at $z=0$ calculated from the NLFFF extrapolation. (b)--(i) Maps of $Q$ slices in the $x$-$z$ planes with different $y$ values. The red star in panel (i) marks the position of the null-point. L1 and L2 correspond to the cuts in the $x-z$ plane of the western and the eastern structures (S1 and S2) of the large-scale QSL induced by the quadrupolar magnetic field.}
\label{fig:7}
\end{figure}

\begin{figure}
\centering
\includegraphics[width=1\textwidth]{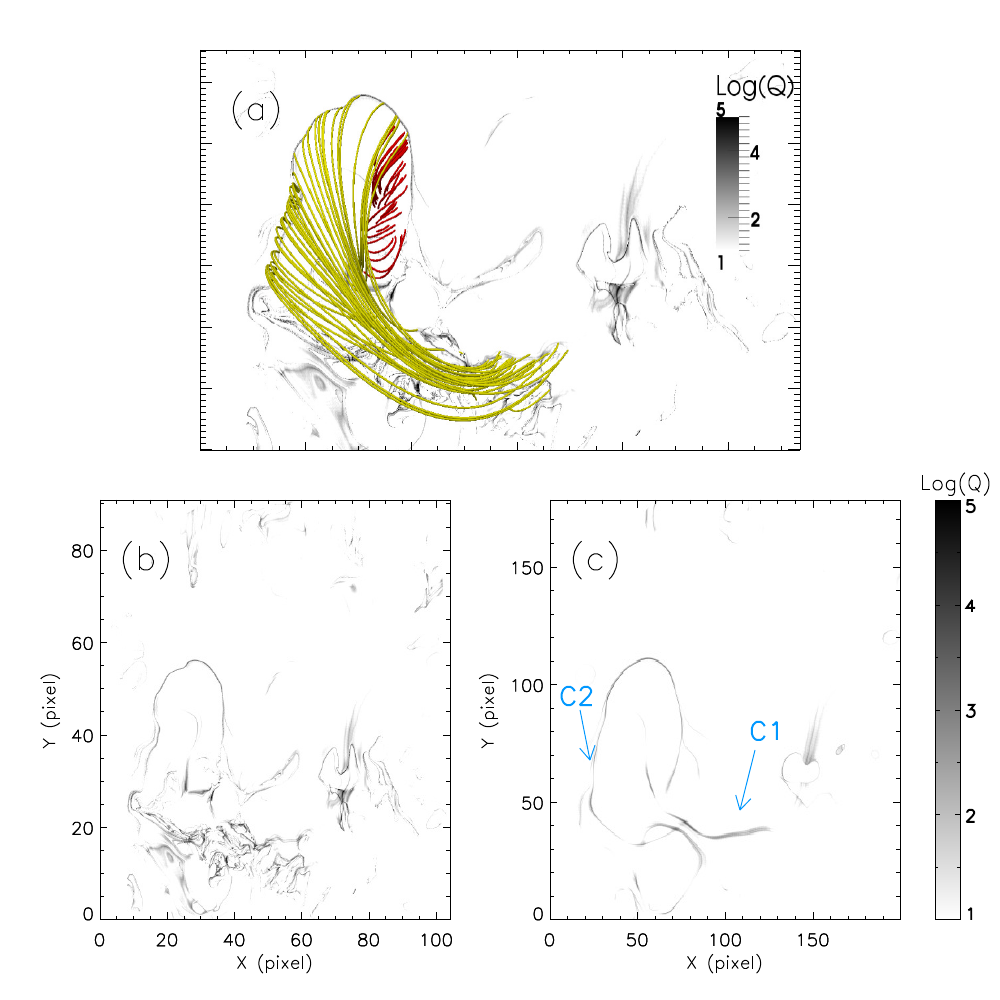}
\caption{(a) Magnetic field after the flare calculated from the NLFFF extrapolation, in which the red field lines are highly sheared and the yellow field lines belong to the QSL. (b)--(c) Maps of squashing factor $Q$ on the bottom boundary calculated from the NLFFF extrapolation and the potential field, respectively. The double C-shaped footprints of the large-scale QSLs after the flare are labelled as C1 and C2, respectively.}
\label{fig:8}
\end{figure}

After the above comprehensive analysis, we come up with a plausible interpretation for the emission features and time evolution of this flare, which involves three steps.
First, a flux rope already lies under the fan structure before the onset of the flare; then, magnetic reconnection occurs inside or near the border of the large flux rope, although we cannot determine where it takes place accurately.
The rising flux rope, appearing as a large and bright ALS (Figure \ref{fig:3}), may further incur magnetic reconnection with the surrounding arcade field lines.
This process feeds new flux to the flux rope and makes it growing gradually.
Meanwhile, the electrons are accelerated and spiral downward along the field lines to the feet of the flux rope, producing two HXR footpoint sources through thick target bremsstrahlung, similarly to the event studied by \citet{2012Guo}. 
However, we cannot make it clear what causes the time difference of the two HXR sources, which may probably be caused by the variation in magnetic and thermodynamic structures along the flux rope.
With the flux rope expansion and the reconnection going on, the flux rope footpoints likely experience a slight movement parallel to the polarity inversion line that may account for the southward movement of the HXR source in the 12--25 keV band (Figure \ref{fig:2}).
Second, the flux rope rising may induce the formation of a current sheet just below it where the reconnection results in the two flare ribbons in \ion{Ca}{2} H, as is predicted in a 3D extension of the CSHKP model proposed by \cite{2012Aulanier}.
At the initial time, the two ribbons are very close to each other appearing just like one ribbon, but they can be distinguished in the movie attached to Figure \ref{fig:1}.
Then, they separate from each other and the western one evolves to a quasi-circular one.
Their locations trace the inner QSL and the western part of the fan, respectively.
Therefore, in the lightcurves, the central ribbon rises first while the quasi-circular ribbon starts to rise about half a minute later.
Third, the rising flux rope pushes the fan structure and the large-scale QSL above and drives the reconnection between the field lines embedded in the QSL and fan.
This process is supported by the time sequence of AIA 94 \AA \ observations (Figures \ref{fig:3}j--l), which shows a dome-like structure being squeezed by the rising flux rope.
As a consequence of reconnection, electrons are accelerated and transported downward along the field lines in the QSLs to heat the chromospheric plasma.
Then, the \ion{Ca}{2} H emission displays a morphology similar to the QSLs at the bottom boundary.
It is seen that the rising flux rope not only squeezes the structure S2 of the large-scale QSL that corresponds to the dome-like structure but also drags the structure S1 of the large-scale QSL that corresponds to the loop-like structure in AIA 94 \AA \  (Figure \ref{fig:3}l).
Such a dragging process induces magnetic reconnection in the large-scale QSL and thus the remote ribbon appears to be located just at the western footprints of this QSL.
The remote ribbon is thought to be heated by energetic electrons transported through the large-scale field lines in the large-scale QSL; therefore, it appears some time later than the two ribbons mentioned above.
After the flare, the magnetic field structure contains no null-point, which suggests that the magnetic reconnection destroys it during the flare.

Nevertheless, we are not sure which mechanisms work here to trigger this event.
In view of the magnetic reconnection process, this event shows some features consistent partly with the tether-cutting model \citep{2001Moore} and partly with the breakout model \citep{1999Antiochos}, but neither can fully explain the event.
For simplicity, we can define the essence of the tether-cutting model as the core field reconnection (the reconnection inside or near the border of the flux rope, e.g., the reconnection in a possible hyperbolic flux tube under the flux rope) and that of the breakout model as the null-point reconnection (including the QSL reconnection) above the flux rope.
Usually, both the two models are possible to work in one particular event. 
In our case, the analysis indicates a joint  process that involves the flux rope rising, core field reconnection (as in the tether-cutting model), and the null-point reconnection above the flux rope (as in the breakout model).
However, we cannot figure out whether the flux rope rising triggers the reconnection or on the contrary.
What we propose is that the rising of the flux rope and the reconnection process could give a positive feedback to each other. 
While the magnetic reconnection facilitates the rising by increasing the magnetic flux of the flux rope (core field reconnection) and decreasing the magnetic tension force above the flux rope (null-point and QSL reconnection), the flux rope rising incurs an environment to speed up the magnetic reconnection.

Note that in this event, the extrapolated field lines are not cospatial with the ribbons perfectly (Figure \ref{fig:5}b).  
This can be caused by many reasons such as the projection effect, systematic errors of the extrapolation, and the non-force-freeness of photospheric field.
Moreover, there also exists an uncertainty in the location of the HXR sources owing to the limited spatial resolution of the RHESSI reconstructed images.  
Therefore, at present we cannot give more details related to the flare process.
Observations with higher spatial and temporal resolution would be required in the future in order to make a better understanding of such events containing 3D null-point reconnection.

\acknowledgments
The authors thank Haimin Wang for valuable discussions and suggestions.
Thanks also go to the referee for many constructive comments that led to a significant improvement of the paper.
This work was supported by NSFC under grants 11373023 and 11203014, and NKBRSF under grants 2011CB811402 and 2014CB744203.



\end{document}